\documentclass[aps,twocolumn,showpacs,amsmath,amssymb,superscriptaddress,final,prb,a4paper,10pt]{revtex4-1}
\usepackage[latin1]{inputenc}   
\usepackage{dcolumn}  
\usepackage{nicefrac}   
\usepackage[dvips]{color} 

\usepackage[charter,greekuppercase=italicized]{mathdesign}\usepackage{sidecap}

\definecolor{red}{rgb}{0.85,.1,0}
\definecolor{cblue}{named}{CadetBlue}
\usepackage{graphicx}  
\graphicspath{{./}{figures/}}
\DeclareGraphicsExtensions{.eps,.png,.pdf}

\newcommand{\nfao}{Nd\-Fe\-As\-O$_{1-x}$\-F$_{x}$}

\newcommand{\musr}{$\mu$SR}

\begin{document}
\title{Slow magnetic fluctuations and superconductivity in fluorine-doped NdFeAsO}

\author{G.~Lamura}
\affiliation{CNR-SPIN and Universit\`a di Genova, via Dodecaneso 33, I-16146 Genova, Italy}
\author{T.~Shiroka}
\email[Corresponding author: ]{tshiroka@phys.ethz.ch}
\affiliation{Laboratorium f\"ur Festk\"orperphysik, ETH-H\"onggerberg, CH-8093 Z\"urich, Switzerland}
\affiliation{Paul Scherrer Institut, CH-5232 Villigen PSI, Switzerland}

\author{P.~Bonf\`a}
\affiliation{Dipartimento di Fisica e Scienze della Terra, Universit\`a  degli Studi di Parma, Viale G. Usberti 7A, I-43124 Parma, Italy}
\author{S.~Sanna}
\affiliation{Dipartimento di Fisica and Unit\`a CNISM di Pavia, I-27100 Pavia, Italy}
\author{R.~{De~Renzi}}
\affiliation{Dipartimento di Fisica e Scienze della Terra, Universit\`a  degli Studi di Parma, Viale G. Usberti 7A, I-43124 Parma, Italy}
\author{M.~Putti}
\affiliation{CNR-SPIN and Universit\`a di Genova, via Dodecaneso 33, I-16146 Genova, Italy}
\author{N.~D.~Zhigadlo}
\affiliation{Laboratorium f\"ur Festk\"orperphysik, ETH-H\"onggerberg, CH-8093 Z\"urich, Switzerland}
\author{S.~Katrych}
\affiliation{Laboratoire de Physique de la Mati\`ere Complexe, EPFL, CH-1015 Lausanne, Switzerland}
\author{R.~Khasanov}
\affiliation{Paul Scherrer Institut, CH-5232 Villigen PSI, Switzerland}
%
%
\author{J.~Karpinski}
\affiliation{Laboratoire de Physique de la Mati\`ere Complexe, EPFL, CH-1015 Lausanne, Switzerland}
\date{\today}

\begin{abstract}

Among the widely studied superconducting iron-pnictide compounds belonging to the Ln1111 family (with Ln a lanthanide), a systematic investigation of the crossover region between the superconducting and the antiferromagnetic phase for the Ln\,$=$\,Nd case has been
missing. We fill this gap by focusing on the intermediate doping regime of Nd\-Fe\-As\-O$_{1-x}$\-F$_{x}$ by means of dc-magnetometry and muon-spin spectroscopy ($\mu$SR) measurements. The long-range order we detect at low fluorine doping is replaced by short-range magnetic interactions at $x = 0.08$, where also superconductivity appears.
In this case, longitudinal-field $\mu$SR experiments show clear evidence of slow magnetic fluctuations that disappear at low temperatures. This fluctuating component is ascribed to the glassy-like character of the magnetically ordered phase of NdFeAsO at intermediate fluorine doping.
\end{abstract}

\maketitle
\section{\label{sec:intro}Introduction}
The simultaneous presence of superconductivity (SC) and magnetism (M), competing and/or coexisting in various families of iron-based superconductors, has captured the interest of the condensed matter community for many years now. Particular attention has been devoted to the Ln1111 family (Ln being a lanthanide), where the static long-range antiferromagnetic order (AFM), occurring in the FeAs planes, is progressively suppressed in favor of the superconducting state by carrier-doping of the parent compound. In the Ln1111 family the M-SC crossover region is characterized by the coexistence of a short-range AFM order and superconductivity on a nanoscopic length scale. This distinctive feature has been shown
to be independent of the doping agent, since it occurs both in H-doped La1111,\cite{Lamura2014} as well as on F-doped Ce-1111,\cite{Shiroka2011}
or Sm-1111.\cite{Sanna2009} While this suggests that H or F act merely as electron donors (although with different efficiencies), it leaves open the question of the lanthanide ions, i.e.: are the M-SC ground-state properties of the Ln1111 family also independent of the Ln ion?

To address this issue and to better understand the role of LnO planes on the occurrence of a M/SC ground state, we considered the case of \nfao, a family about which relatively fewer studies have been published insofar, none of which addressing this important point. In particular, since the discovery of superconductivity in a Nd1111 compound in 2008,\cite{RenEu2008} most of the studies focused on the evolution of the structural and superconducting properties induced either by F-doping,\cite{Malavasi2010} or by simple oxygen deficiency.\cite{Ren2008,Lee2008} The magnetic properties of the undoped compound and those of an $x(\mathrm{F})=0.06$ under-doped sample were reported in Refs.~\onlinecite{McGuire2009} and \onlinecite{Tarantini2008}, respectively. In the latter case it was emphasized that the dc magnetization response is dominated by the Curie-Weiss paramagnetism of the Nd$^{3+}$ ions, whereas no signs of long-range antiferromagnetic order, due either to Nd$^{3+}$ or to
Fe$^{2+}$ ions, could be detected below $T_
c$. Later on,
 M\"{o}ssbauer-effect studies confirmed the absence of magnetically-ordered phases in an $x(\mathrm{F})=0.12$ superconducting sample.\cite{Baggio2009}
\musr\ measurements on either undoped,\cite{JPCarlo2009} or on F-doped\cite{Aczel2008} samples were performed, too. In the former case they could detect an AFM long-range magnetic order with $T_\mathrm{N}=135$\,K and a local field $B_\mu=0.17$\,T, while in the latter a gapped-like superfluid density was inferred. Unfortunately, due to the lack of good-quality, finely-tuned F-doped samples, a systematic study of the most interesting region, that of the M-SC crossover, has been missing to date.\\
In the present paper we address exactly this issue by investigating via \musr\ four different F-doped polycrystalline samples with $x(\mathrm{F})=0.03$, 0.05, 0.08, and 0.25. In the first two cases we find a long-range AF order, that gradually extinguishes as the F content increases. On the other hand, in the $x(\mathrm{F})=0.25$ sample, we do not find any  magnetically ordered phases. The intermediate $x(\mathrm{F})=0.08$ case, instead, is particularly remarkable since here we detect the simultaneous presence of bulk superconductivity (below $T_c=20$\,K) and of short-range magnetic order (below  $T_{\mathrm{N}}=40$\,K). Interestingly, in an intermediate
$T_c < T < T_{\mathrm{N}}$ temperature range, sizable spin-fluctuations with a slow dynamics dominate. We ascribe these magnetic spin fluctuations to the glassy character of the ordered magnetic phase.\\
In a broader context, the current study of the Nd1111 family fills a gap about the role played by the different rare earths in determining the properties of Ln1111 compounds, therefore allowing detailed comparisons to be made.

\section{\label{sec:exp_details}Experimental}
\subsection{\label{ssec:structural}Sample synthesis and structural characterization}
A series of \nfao\ polycrystalline samples, with nominal fluorine content ranging from $x=0.03$ to $x=0.25$,
was prepared at ETH Zurich using cubic-anvil, high-pressure, and high-temperature techniques.\cite{Khasanov2011,Zhigadlo2008,Zhigadlo2010}
Due to the toxicity of arsenic, all procedures related to the sample preparation were performed in a glove box.
Pellets containing the high-purity ($\geq 99.95\%$) precursors (NdAs, FeAs, Fe$_2$O$_3$, Fe, and NdF$_3$)
were enclosed in a boron nitride container and placed inside a graphite heater. Six tungsten carbide anvils generated
the required high pressure (3 GPa) on the assembly. While the pressure was kept constant, the temperature was ramped up in 2 h, from $20^{\circ}$C  to the
maximum value of $\sim 1350^{\circ}$C, and maintained there for 15 h. Following the temperature quenching, the pressure was released and the sample removed.

\begin{figure}[tbh]
\centering
\includegraphics[width=0.42\textwidth]{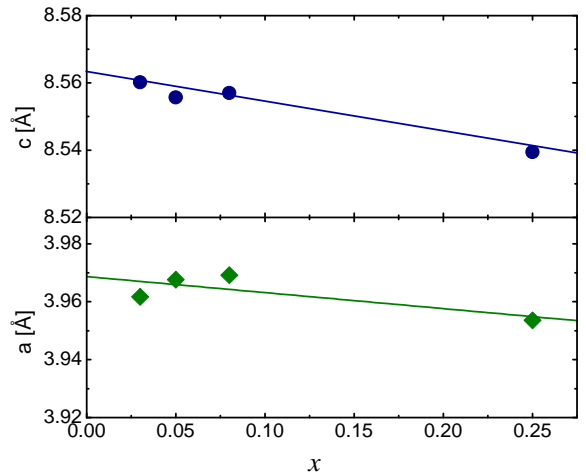}  
\caption{\label{fig:fig1}(color online). Evolution of the $a-$ and $c-$axis lengths of \nfao\ at 300 K as a function of the nominal fluorine content. Continuous lines are only guides to the eye.}
\end{figure}

The structural characterization was performed by means of standard powder x-ray diffraction (XRD) measurements carried out at room temperature. The crystalline structure was refined using the Rietveld technique. The lattice constants, the type of impurities, and the respective volume fractions are reported in Table~\ref{tab:xrdiff}.
In particular, the variations of the unit-cell parameters as a function of the nominal fluorine content are plotted in Fig.~\ref{fig:fig1}. We note two important features: (\textit{i}) the $a$-axis length is slightly reduced with doping. This effect is almost Ln-independent and has been ascribed to the presence of small oxygen deficiencies in samples grown at high pressure.\cite{Zhigadlo2008,Kito2008,Lee2008} (\textit{ii}) The $c$-axis length decreases linearly with increasing $x$(F), too. In this case, the replacement of oxygen with the more electronegative fluorine implies a reduction of the unit-cell volume as the F content increases. This result is similar to that observed also in other F-doped 1111 series (see, e.g., Ref.~\onlinecite{Shiroka2011} for the Ce case) and demonstrates the successful F-to-O substitution.
\subsection{\label{ssec:dc_squid}Magnetometry measurements}
DC magnetization measurements were performed by means of a superconducting quantum interference device (SQUID) magnetometer (Quantum Design) on all the tested samples. Both magnetization vs.\ temperature (from 2 up to 300 K) and isothermal magnetization measurements vs.\ applied field, $m(H)$, at selected temperatures were carried out. The linear (i.e.\ purely paramagnetic) behavior of $m(H)$ at both low and high temperatures (not shown) demonstrates the absence of diluted ferromagnetic impurities. This was confirmed also by $\mu$SR measurements at high temperature (see below), where no
significant asymmetry damping could be observed within the 5~$\mu$s data-acquisition window used.

\begin{figure}[tbh!]
\centering
\includegraphics[width=0.42\textwidth]{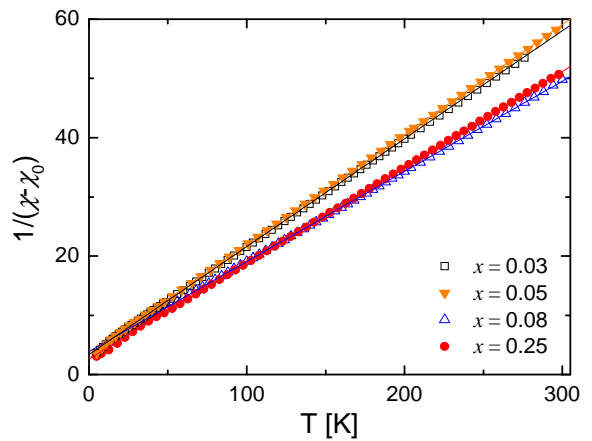}  
\caption{\label{fig:fig2}(color online). Temperature dependence of the inverse dc susceptibility in \nfao. Data were recorded in zero-field cooled (ZFC) mode in an applied field $\mu_0 H = 3$ T. Lines represent fits using the $1/[\chi(T)-\chi_0]$ model for $T>50$\,K (for consistency among the normal and superconducting samples).}
\end{figure}

In Fig.~\ref{fig:fig2} the zero-field cooled dc magnetization measurements carried out at $\mu_0 H = 3$ T are plotted as inverse susceptibility vs.\ temperature. Numerical fits of $\chi(T)$ data with a Curie-Weiss law restricted to the high-temperature regime allowed us to determine $\chi_0$, the residual temperature-independent contribution to the susceptibility. Successively, a linear fit of the data with $1/[\chi(T)-\chi_0]$ provided the Curie constant $C$, from which the Nd$^{3+}$ magnetic moment $\mu_{\mathrm{eff}}$ (in the free-ion approximation) could be extracted. The obtained values lie in the range  $\mu_{\mathrm{eff}} = 3.5$--3.8\,$\mu_{\mathrm{B}}$, i.e., very close to the value expected for the Nd$^{3+}$ free-ion magnetic moment (3.62 $\mu_{\mathrm{B}}$).
No evidence of Nd antiferromagnetic order was found down to 2 K, in agreement with previous neutron diffraction studies,\cite{Qiu2008,Chatterji2011} which show that a possible AF ordering of Nd moments most likely occurs below 2 K.

\begin{figure}[tbh]
\centering
\includegraphics[width=0.4\textwidth]{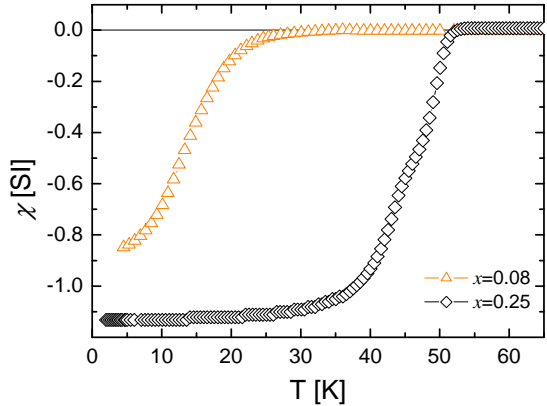}  
\caption{\label{fig:fig3}(color online). Temperature dependence of the
volume magnetic susceptibility in \nfao\ for $x = 0.08$ and 0.25 in an applied field of 1 mT, following zero-field cooling.}
\end{figure}

At low temperature, the samples with higher fluorine doping, $x = 0.08$ and 0.25, display a clear superconducting transition, as shown by the low-field (1 mT) ZFC magnetization data in Fig.~\ref{fig:fig3} and summarized in Table~\ref{tab:TCTN}. The superconducting shielding fraction\footnote{It was calculated by using the nominal density and no demagnetizing factor, since the applied field was set parallel to the main surface of the disc-shaped samples.}  of about 80\% is most likely a lower bound, since at low doping values the magnetic penetration depth $\lambda_L$ becomes comparable to the grain size (micrometer scale), hence resulting in a reduced Meissner response. This result  demonstrates the bulk character of the superconductivity in these samples.
\begin{table*}[thb]
\centering
\renewcommand{\arraystretch}{1.2}
\caption{\label{tab:xrdiff} Unit cell parameters and impurity fractions in \nfao\ samples as determined by Rietveld refinement of x-ray diffraction data taken at room temperature.}
\begin{ruledtabular}
\begin{tabular}{p{1mm}lllccccc}
\vspace{2mm}
& \lower 0.7mm \hbox{$x$(F)}& \lower 0.7mm \hbox{$a$-axis (\AA)}& \lower 0.7mm \hbox{$c$-axis (\AA)}& \lower 0.7mm \hbox{Pure phase (\%)}& \lower 0.7mm \hbox{FeAs (\%)}& \lower 0.7mm \hbox{NdOF (\%)}& \lower 0.7mm \hbox{NdAs (\%)}& \lower 0.7mm \hbox{Other (\%)}\\[2pt]
\hline
&0.03 &3.9669(2)&8.5602(5)&98(2)& --    & --     & --    & $<2$  \\
&0.05 &3.9676(2)&8.5557(5)&98(2)& --    & --     & --    & $<2$  \\
&0.08 &3.9617(4)&8.5570(10) &93(3)&4.11(9)& 2.3(1) &0.38(5)&0.21(1)\\
&0.25 &3.9536(1)&8.5394(5)&84(1)& --    &9.8(0.4)&5.7(2) & 0.5(1)\\
\end{tabular}
\end{ruledtabular}
\end{table*}
\subsection{\label{ssec:muSR}Muon-spin spectroscopy}
Muon-spin relaxation measurements were carried out at the GPS spectrometer ($\pi$M3 beam line) at Paul Scherrer Institut (PSI), Villigen, Switzerland. 100\% spin-polarized muons were implanted uniformly over the sample volume and successively the relevant decay positrons were detected.
Once muons thermalize at interstitial sites (almost instantaneously and without loss of polarization), they act as sensitive probes, precessing in the local magnetic field $B_{\mu}$ with a frequency $f_{\mu} = \gamma_{\mu}/(2\pi)\, B_{\mu}$, where $\gamma_{\mu}/(2\pi) = 135.53$ MHz/T is the muon gyromagnetic ratio. Both zero-field (ZF) and longitudinal-field (LF) $\mu$SR experiments were performed. Due to the absence of externally applied fields, ZF-$\mu$SR represents the best technique for investigating the spontaneous magnetism and its evolution with fluorine doping. On the other hand, LF-$\mu$SR measurements can be successfully used to single out the dynamic or static character of the internal magnetic fields.\cite{Yaouanc2011}

\subsubsection{\label{ssec:ZFmuSR}Zero-field $\mu$SR measurements}
\begin{figure}[tbh]
\centering
\includegraphics[width=0.42\textwidth]{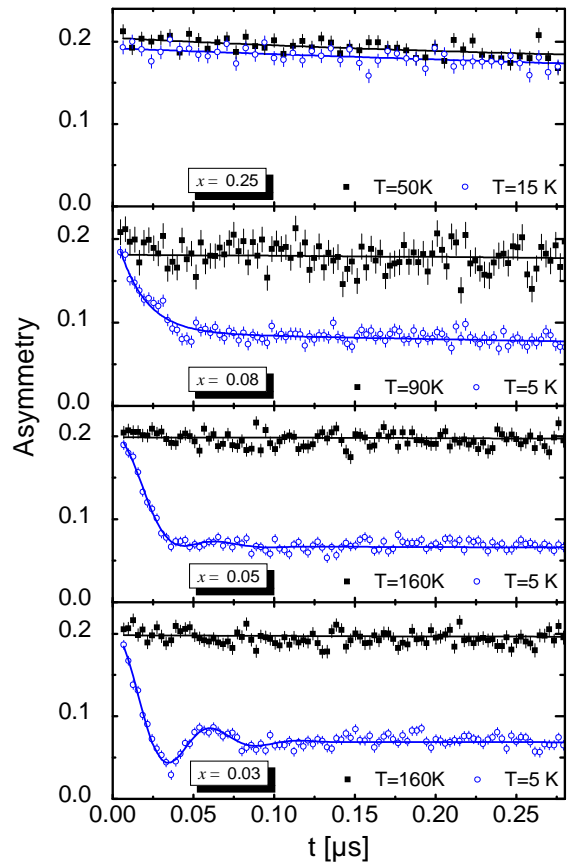} 
\caption{\label{fig:fig4}(color online). ZF-$\mu$SR asymmetry data for \nfao\ samples, collected at high and low temperatures. Solid lines represent numerical fits by means of Eqs.~(\ref{eq:osc}) and (\ref{eq:osc_008}). See text for details.}
\end{figure}
Figure~\ref{fig:fig4} shows the short-time ZF asymmetry (i.e.\ the  muon-spin precession signal), $A(t)$, for all the tested \nfao\ samples at both low and high temperatures. At high temperatures, i.e.\ above the N\'eel temperature $T_{\mathrm{N}}$, the asymmetry signal is practically flat, with no oscillations and with a negligible decay of the initial polarization, mainly due to the randomly-oriented nuclear magnetic moments. When the temperature is lowered below $T_{\mathrm{N}}$, the asymmetry becomes highly damped, reflecting the  presence of a spontaneous magnetic order whose character depends on $x$(F): at low fluorine content we observe well-defined asymmetry oscillations which, as $x$(F) increases, change quickly to highly damped ones. This behavior is indicative of a \emph{long-range} (static) uniform magnetic order. For $x = 0.08$ the damping becomes so high that the oscillatory behavior disappears altogether, to be replaced by a fast decaying signal, typical of a wide distribution of fields
attributed to \emph{short-range} magnetic order. Finally, in the $x = 0.25$ case, we find no difference between the low- and high-temperature asymmetries, which implies the absence of a magnetically ordered phase.\footnote{In the $x=0.25$ case we plot the asymmetry at 15 K, since a small magnetic component ($\sim$ 10\% of the total asymmetry) appears below 5 K, most likely due to impurity phases. This fraction is compatible with x-ray diffraction analysis summarized in Table~\ref{tab:xrdiff}. Since the weight of the impurity phases is very low, we neglect it in the following discussion.}
To go beyond the above qualitative results and achieve a better understanding, the various ZF-$\mu$SR time-domain data were fitted using the function:
\begin{equation}
\label{eq:osc}
\frac{A^{\mathrm{ZF}}(t)}{A_{\mathrm{tot}}^{\mathrm{ZF}}(0)}= \left[ a^{A}_{T} e^{-\frac{\sigma_A^2 t^2}{2}} f (\gamma B_{\mu} t) \!
+ a^{B}_{T} e^{-\frac{\sigma_B^2 t^2}{2}}\right]+ a_{L}e^{-\lambda_L t},
\end{equation}
where $a^{A,B}_{T}$ are the two transverse asymmetry components, while $a_{L}$ represents a longitudinal component, each with respective decay coefficients $\sigma_{A,B}$ and $\lambda_L$. The transverse components refer to the two distinct muon implantation sites, normally expected for the 1111 family: the most populated site, named A,  located next to the FeAs planes, and a secondary one, named B, close to the oxygen atoms in the LnO planes.\cite{Maeter2009} Since only $\sim 15\%$ of the total implanted muons are located in B sites, their contribution to the total asymmetry is too small to allow the extraction of a second precession frequency.\cite{Shiroka2011} Therefore, we describe it only by a Gaussian decay ($a^{B}_{T}$). The corresponding longitudinal components for muons stopped in both sites share similar decay rates and, hence, can be merged into a single term, $a_{L}$.

The analytical form of the term $f(t)$ depends on the F content: for $x=0.03$ and 0.05 the best fit was obtained with $f(t)=J_0(\gamma B_{\mu} t)$, the zeroth order Bessel function, generally attributed to \textit{incommensurate} long-range ordered systems.\cite{Yaouanc2011} On the other hand, for $x=0.08$, no coherent oscillations could be detected [$f(t)=1$] and the time-dependent asymmetry was better fitted using the expression:
\begin{equation}
\label{eq:osc_008}
\frac{A^{\mathrm{ZF}}(t)}{A_{\mathrm{tot}}^{\mathrm{ZF}}(0)}= \left[ a^{A}_{T} e^{-\lambda_A t} \!
+ a^{B}_{T} e^{-\lambda_B t}\right]+ a_{L}e^{-(\lambda_L t)^\beta}.
\end{equation}
The transverse component $a_T(t)$ was fitted by a simple exponential decay, whereas for the longitudinal one, $a_L(t)$, a stretched exponential was chosen in order to distinguish possible dynamic contributions (\textit{vide infra}), with $\beta$ the stretching coefficient.
\begin{figure}[tbh]
\centering
\includegraphics[width=0.42\textwidth]{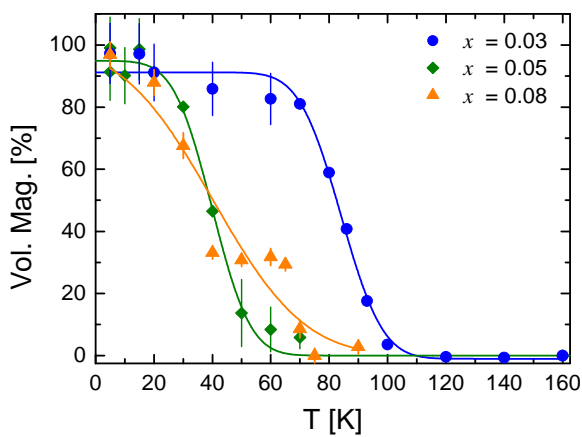} 
\caption{\label{fig:fig5}(color online). Temperature dependence of the magnetic
volume fraction for samples with $x(\mathrm{F}) =$ 0.03, 0.05, and 0.08. Solid lines represent numerical fits by means of Eq.~(\ref{eq:mag_vol_frac}).}
\end{figure}
All the considered samples were po\-ly\-cry\-stal\-li\-ne, therefore, the randomly-oriented internal-field model was invoked to calculate the low-temperature magnetic volume fraction: when the whole sample shows static magnetic order, on average $\nicefrac{1}{3}$ of the muons experience a field parallel to their initial spin direction and do not precess (longitudinal component), while the remaining  $\nicefrac{2}{3}$ precess around the local field (transverse component). This is exactly what we find at low temperature for the samples $x = 0.03$, 0.05, and 0.08 (see Fig.~\ref{fig:fig4}). Therefore, all these samples can be considered \emph{fully} magnetically ordered at low temperature.
To track the temperature dependence of the magnetic volume fraction, $V_m$, the following procedure was used. From the longitudinal component, as  extracted from Eq.~(\ref{eq:osc}), one can calculate $V_m(T) = \frac{3}{2} \left(1-a_{L}\right) \cdot 100\%$. This is the case of the sample $x = 0.08$ and, for temperatures below $\sim 10$ K, also of the samples $x = 0.03$ and 0.05. For the latter two, $V_m$ at higher temperatures could be extracted using the expression $V_m(T) = \left(1-a_{\mathrm{para}}\right) \cdot 100\%$, where $a_{\mathrm{para}}$ represents the muon asymmetry related to paramagnetic (i.e., nonmagnetic) species, as resulting from a transverse-field (TF) $\mu$SR experiment,
typically performed at 3 mT. The obtained $V_m(T)$ data for $x = 0.03$, 0.05, and 0.08 cases are shown in Fig.~\ref{fig:fig5}. To determine the average N\'eel temperatures and the corresponding transition widths, the $V_m(T)$ data were fitted using the following phenomenological function:
\begin{equation}
\label{eq:mag_vol_frac}
V_m(T) = \frac{1}{2} \left[ 1 -\mathrm{erf} \left(\frac{T-T_{\mathrm{N}}}{\sqrt{2}\Delta_{T_N}}\right)\right].
\end{equation}
Here erf is the cumulative error function (typical of a normal distribution). The fit results are summarized in Table~\ref{tab:TCTN}. In general, as the fluorine content $x$(F) increases there is both a gradual decrease of the average $T_{\mathrm{N}}$ and a progressive broadening of the transition.

\begin{figure}[tbh]
\centering
\includegraphics[width=0.40\textwidth]{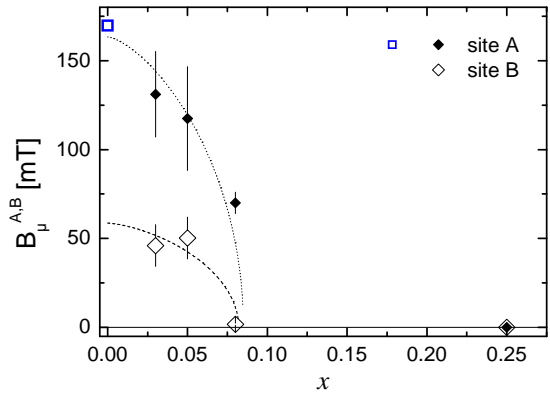} 
\caption{\label{fig:fig6}(color online). Doping dependence of the internal field values as probed by implanted muons in sites A (filled symbols) and B (open symbols) at 5 K. For the low-doped samples, $x = 0.03$ and 0.05, both the local-field value $B_\mu^A$ at site A and its width $\Delta B_{\mu}^A=\sigma_A / \gamma_\mu$ (plotted as error bar) could be determined, whereas for the site B the field distribution width $\Delta B_{\mu}^{B}=\sigma_B / \gamma_\mu$ is shown. For the $x = 0.08$ sample, only
$\Delta B_{\mu}^{A,B}=\lambda_{A,B} / \gamma_\mu$ are displayed. The open square represents the internal field at site A for the undoped compound,  as reported in Ref.~\onlinecite{Aczel2008}. The dotted and dashed curves are guides to the eye.}
\end{figure}
Fits of low-temperature asymmetry data with Eq.~(\ref{eq:osc}) allow us to determine
also the internal field values at the lowest temperature, as probed by muons implanted in sites A and B. The dependence of these fields on fluorine doping $x$(F) is shown in Fig.~\ref{fig:fig6}. As the fluorine content increases, the average field value $\left< B_{\mu} \right>$
decreases and its Gaussian distribution broadens. The $x = 0.08$ case deserves particular attention: $A(t)$ does not show any coherent oscillation and the internal field can be modeled by a broadened Lorentzian distribution of fields, whose values range from zero up to $\lambda_{A,B}/\gamma_\mu$.

\begin{figure}
\centering
\includegraphics[width=0.42\textwidth]{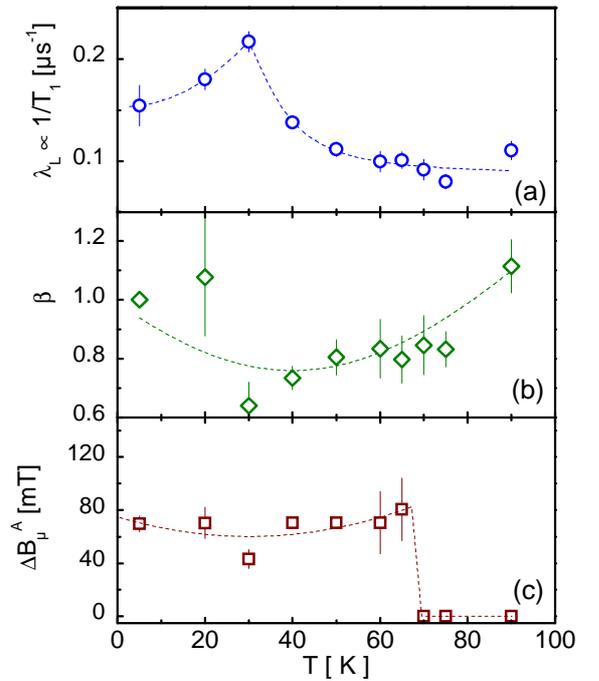}
\caption{\label{fig:fig7} (color online). $x = 0.08$ T-dependence of the parameters extracted by fitting the time-domain $\mu$SR asymmetry data by means of Eq.~(\ref{eq:osc_008}). The dashed curves are guides to the eye.}
\end{figure}

In Fig.~\ref{fig:fig7} we show the temperature behavior of the main parameters obtained by fitting the time-dependent asymmetry by Eq.~(\ref{eq:osc_008}). We notice that: (a) the longitudinal relaxation rate has a peak at about 30 K; (b) the stretching coefficient shows a broad minimum centered again at 30 K; (c) the field width probed by the implanted muons at site A is almost constant up to the onset of the magnetic order, while the field probed at site B is very low (2--5 mT) and almost undetectable above 30 K (not shown).
\begin{table}[hbt]
\centering
\renewcommand{\arraystretch}{1.2}
\caption{\label{tab:TCTN} Magnetic properties of the \nfao\ family, as determined from $\mu$SR and dc magnetometry measurements. The $T_{\mathrm{N}}$ of the undoped compound is taken from Ref.~\onlinecite{Aczel2008}.}
\begin{ruledtabular}
\begin{tabular}{p{1mm}lccccc}
\vspace{2mm}
& \lower 0.4mm \hbox{$x$(F)}
& \lower 0.4mm \hbox{$T_{\mathrm{N}}$~(K)}
& \lower 0.4mm \hbox{$\Delta T_{\mathrm{N}}$~(K)}
& \lower 0.4mm \hbox{$V_{\mathrm{mag}}$~(\%)}
& \lower 0.4mm \hbox{$T_{c}$(K)}\\[2pt]
\hline
&0    & 135   &         &        &     \\
&0.03 &  84(1)&  11(1)  &  $>$90 & --  \\
&0.05 &  40(1)&  10(1)  &  $>$90 & --  \\
&0.08 &  40(3)&  25(4)  &  $>$90 &20(2)\\
&0.25 &   --  &   --    &  --    &51(1)\\
\end{tabular}
\end{ruledtabular}
\end{table}
\subsubsection{\label{ssec:LFmuSR}Longitudinal-field $\mu$SR measurements}
The sample $x=0.08$ is the most interesting one, since, at low temperature, bulk superconductivity coexists with an ordered magnetic state. The absence of muon-spin precession and a large damping of asymmetry at short times clearly indicate the presence of a wide distribution of fields.
However, it is not obvious whether this distribution is due to static fields, or to strongly fluctuating (i.e., dynamic) magnetic moments. To disentangle these two possibilities, we carried out longitudinal-field (LF) $\mu$SR measurements at 30\,K. This temperature represents  a good compromise, since it is above the superconducting critical temperature $T_c \simeq 20$ K (see Fig.~\ref{fig:fig3}), yet sufficiently low to  admit an ordered volume fraction of about 70\% (see Fig.~\ref{fig:fig5}). In an LF-decoupling experiment\cite{Hayano1979} an external magnetic field $B_{\parallel}$ is applied along the initial muon-spin direction. If $B_{\parallel}$ is of the same order as or higher than the internal static fields, then it will significantly affect the muon polarization through a ``spin-locking'' effect. On the other hand, in case of strongly fluctuating internal fields, the effect of the external field is barely noticeable.

\begin{figure}
\centering
\includegraphics[width=0.40\textwidth]{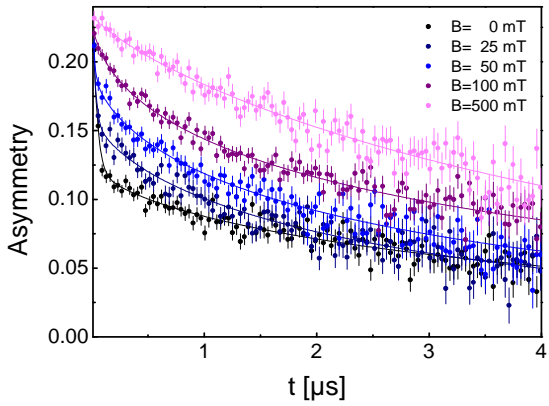}
\caption{\label{fig:fig8} (color online). Time-domain LF-asymmetry data taken at $T = 30$~K in the $x=0.08$ sample for different values of the longitudinal field. Solid lines represent numerical fits by means of Eq.~(\ref{eq:osc_008}).}
\end{figure}

\begin{figure}
\centering
\includegraphics[width=0.42\textwidth]{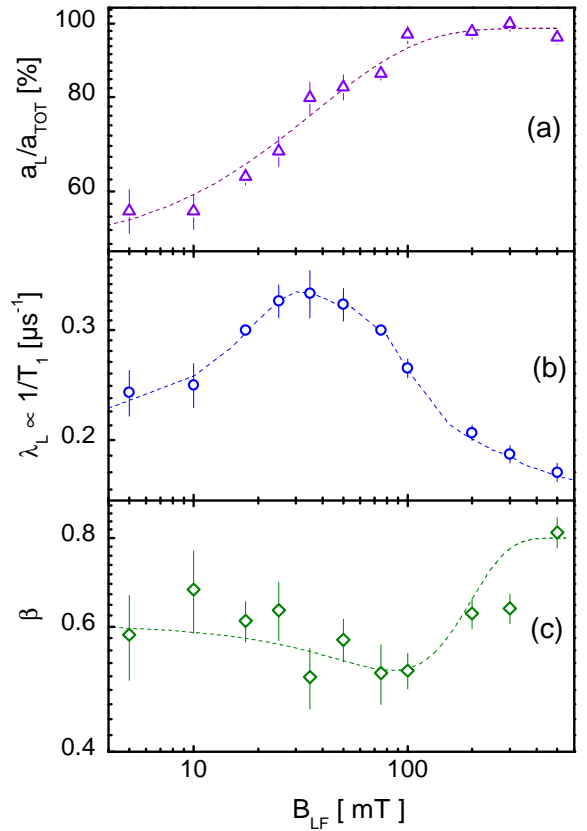}
\caption{\label{fig:fig9} (color online). Longitudinal-field dependence of the parameters extracted by fitting the time-domain asymmetry data taken at $T = 30$~K in the $x=0.08$ sample (see Fig.~\ref{fig:fig8}). The dashed curves are guides to the eye.}
\end{figure}

In Fig.~\ref{fig:fig8} the time-dependent asymmetry is shown for a selection of $B_{\parallel}$ values. The time-domain data were successfully fitted by means of Eq.~(\ref{eq:osc_008}), with the field dependence of the resulting best-fit parameters for the longitudinal component being shown in Fig.~\ref{fig:fig9}.
Several interesting and unexpected features can be noticed: (\textit{i}) From the $T$-dependent analysis we know that the static local field can be as high as $\sim 70$\, mT [see Fig.~\ref{fig:fig7}(c)]. Once this is quenched by an applied field of 75 mT, the longitudinal asymmetry is fully recovered only for fields above 0.2\,T, (\textit{ii}) the longitudinal relaxation rate $\lambda_{L} \propto 1/T_1$ has a peak at about 30--50 mT; (\textit{iii}) the exponent $\beta$ reaches 1 only at high fields.
\section{\label{sec:discussion}Discussion}
An understanding of the F-doping effects in Nd1111 compounds, both regarding their magnetic and superconducting properties,
as well as the possible presence of an M-SC crossover region, is expected to provide a more complete picture of the iron-arsenide superconductors.
The results of our systematic investigation of the structural and magnetic properties of representative compounds of the Nd family are summarized and discussed in detail below.

\subsection{\label{sec:general}Static magnetism and superconductivity}
In Nd1111 the macroscopic magnetic behavior (dc susceptibility) is dominated by the paramagnetic response of Nd$^{3+}$ ions, that display an AF order only at 1.96 K (in the undoped case).\cite{Qiu2008} Contrary to the case of Ce1111,\cite{Shiroka2011} where the AF order occurs at higher (i.e., more accessible) temperatures, the rather low values of the Nd family prevented us from verifying possible correlations between $T_{\mathrm{N}}$(Nd) and F-doping.
Given these difficulties, we focused on the magnetism of the Fe$^{2+}$ ions, crucial in determining most of the properties of the Ln1111 compounds.

Below $T_{\mathrm{N}}$(FeAs) (ca.\ 135\,K), a commensurate, long-range AF order develops in the FeAs planes,
giving rise to a coherent precession of the implanted muon ensemble.\cite{Aczel2008}
However, already small values of F-doping are sufficient to induce a high damping in the oscillating muon polarization. This is exactly the case of our $x(\mathrm{F})=0.03$ and 0.05  samples, where the oscillating term is well described by a zeroth-order Bessel function, normally the fingerprint of an incommensurate long-range magnetic order.\cite{Yaouanc2011,SAVICI} When the F content is further increased, the size of the
AF-ordered domains decreases and the magnetic damping at short times becomes so strong that no oscillations are detected in the time-dependent muon polarization. This situations is typical of a short-range magnetic order\footnote{Providing a realistic length scale for the long-to-short range magnetic order transition of arbitrary systems is not easy, since their correlations reflect different factors, including the magnitude of the magnetic moments, the percentage of the ordered moments per magnetic domain, etc.\cite{SAVICI} Yet, by comparison with known systems, one can still provide upper/lower bounds. For instance, in generic ferromagnets, correlation lengths of ca.\ 10 unit cells are sufficient to induce highly-damped coherent precessions (see, e.g., p.\ 264 in Ref.\ \onlinecite{Yaouanc2011}). Conversely, in antiferromagnetic cuprates, which show similar lamellar structures and magnitudes of magnetic moments as in pnictides, this limit can be as low as 3--4 in-plane
unit-cell lengths \cite{SAVICI}} and this is exactly what we found in the case of the sample with $x(\mathrm{F})=0.08$.

\begin{figure}
\centering
\includegraphics[width=0.42\textwidth]{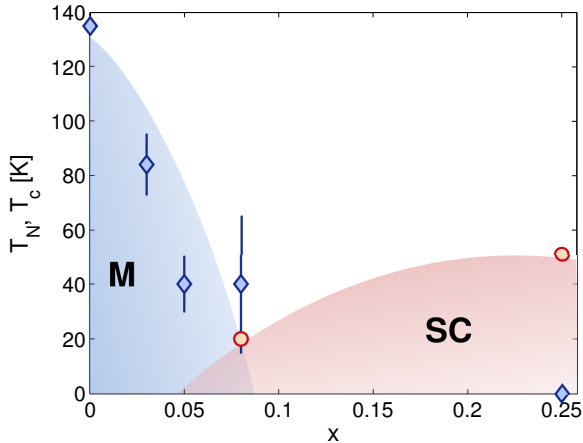} 
\caption{\label{fig:fig10} (color online). Magnetic ordering $T_{\mathrm{N}}$ and critical superconducting temperatures $T_c$ vs.\ fluorine doping $x$ in the \nfao\ family. The lines are guides to the eye. The $T_{\mathrm{N}}$ value for the $x=0$ case is taken from Ref.~\onlinecite{Aczel2008}.}
\end{figure}
By considering the $T_{\mathrm{N}}$ and $T_c$ values inferred by our dc-magnetization and \musr\ data (see Table ~\ref{tab:TCTN}) one can draw a tentative phase diagram that describes the evolution of both M and SC orders as a function of F-doping. As shown in Fig.~\ref{fig:fig10}, the main feature of the phase diagram is the presence of a narrow region where electron doping induces both \textit{bulk} superconductivity and FeAs magnetic order in the \textit{whole} sample volume. The most reasonable way of reconciling the simultaneous presence of M and SC bulk phenomena is to assume their coexistence at a \textit{nanometer length scale}, as already invoked in the case of the Sm1111,\cite{Sanna2009} Ce1111,\cite{Sanna2010,Shiroka2011} and La1111\cite{Lamura2014} families. Therefore, one can most likely conclude that this particular kind of M-SC coexistence is a common feature of all the Ln1111 iron pnictides, independent of the lanthanide ion.

On the other hand, contrary to the above cases, the $x=0.08$ sample of the Nd1111 family shows clear signs of non-negligible \textit{magnetic fluctuations} near the FeAs ordering temperature. This peculiar feature is discussed
in detail in the following section.

\subsection{\label{sec:fluct}Fluctuating magnetism in the $x=0.08$ case}
The muon LF asymmetry data shown in Fig.~\ref{fig:fig8} exhibit two clearly distinct relaxation regimes (see, e.g., data at $B=0$\,mT). At short times, the (fast) relaxation is dominated by static magnetic moments (on the \musr\ time scale), while at long times, the (slow) asymmetry relaxation rate ($\propto 1/T_1$) is mostly determined by dynamic (i.e., fluctuating) magnetic moments. To account for both contributions, the LF-\musr\ data were fitted by means of Eq.~(\ref{eq:osc_008}), where the stretched exponential term can accommodate different situations. Indeed, the stretching coefficient $\beta$ can assume a value $\beta=2$, indicating a Gaussian relaxation process (with the implanted muons probing quasi-static electronic moments with short-range magnetic correlations), or $\beta=1$, indicating a \textit{single} spin-spin autocorrelation time for the entire electronic spin system. The case $\beta<1$, instead, is typical of broad inhomogeneous \textit{distributions} of dynamic (i.e., fluctuating) local
fields (or relaxation rates), which can be due to randomness, disorder, or frustration, as e.g., in the case of spin glasses.

Figures~\ref{fig:fig7}(b) and \ref{fig:fig9}(c), which report the evolution of the $\beta$ parameter with temperature and LF field, respectively, show similar features: $\beta$ exhibits a broad minimum at about $T_{\mathrm{N}}/2$, where it assumes the value 0.6. Then, for temperatures below 20 K, where the magnetic moments in the FeAs planes become static, it grows back again to ca.\ 1. Also $\beta = \beta(B_{\mathrm{LF}})$ shows a broad minimum centered in the 0.05--0.1 T field range, where the longitudinal relaxation rate $\lambda_L$ has a narrow peak reflecting the competition between the static and dynamic components of the local field at the muon site. For applied fields above 75 mT, the $H$-dependence of the longitudinal relaxation rate is no more due to decoupling, but is determined only by the long-lived local field fluctuations. This fact is confirmed by the slow recovery of the longitudinal muon-asymmetry component. A similar behavior was observed also in Tb$_
2$Sn$_2$O$_7$,\cite{Baker2012} where the occurrence of such a peak was attributed to the inapplicability of the Redfield relaxation model and, consequently, reported as evidence of a broad distribution of spin-spin correlation times.

From these observations we argue that in the 20--40\,K temperature interval the local magnetism shows a spin-glass character, most likely due to the proximity of the magnetic transition at $T_{\mathrm{N}}\sim 40$\,K.
Consequently, in this $T$ range one can tentatively use the same approach valid for spin-glass systems. This will allow us to discuss the relation
between the longitudinal-field data and the dynamic spin fluctuations occurring in the magnetic phase.
To this aim we recall that the relaxation at long times probes the spin-fluctuation dynamics and,
in absence of cross-correlations (as is the case for spin glasses), the relaxation rate is just the Fourier transform of the dynamic spin-spin au\-to\-cor\-re\-la\-tion function, $q(t)=\langle \boldsymbol{S}(0) \cdot \boldsymbol{S}(t) \rangle$. When the spin-glass transition is approached from above, $q(t)$ exhibits either a power-law (PL), $q(t) \propto t^{-\alpha}$, or a stretched-exponential (SE) dependence, $q(t) \propto e^{-(\lambda  t )^\beta}$.\cite{Keren1996}
Since in our case the magnetic transition is quite broad (see $x = 0.08$ data in Fig.~\ref{fig:fig5}), we can use the same theoretical approach valid in spin glasses.
The Fourier transform of the au\-to\-cor\-re\-la\-tion function, $\mathfrak{F} \{q(t)\}$, leads then to a time-field scaling of the $\mu$SR asymmetry $A(t,B_{\mathrm{LF}})=A(t/B_{\mathrm{LF}}^\gamma)$, where  $\gamma<1$ for power-law and $\gamma>1$ for stretched-exponential correlations.\cite{Keren1996,Keren2004} This scaling stops being valid at high fields, where $q(t)$ is directly affected by the applied LF field, typically for $\mu_\mathrm{B} B_{\mathrm{LF}} > k_\mathrm{B} T$, a limit which is never reached in our case.

\begin{figure}
\centering
\includegraphics[width=0.42\textwidth]{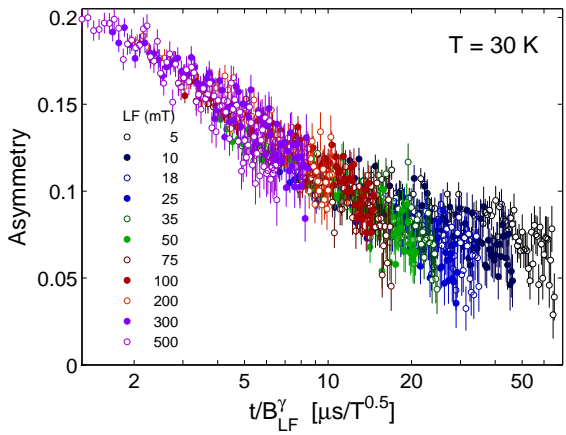} 
\caption{\label{fig:fig11} (color online). Muon asymmetry data taken at $T = 30$~K in the $x=0.08$ sample in various applied longitudinal fields vs.\ the scaled variable $t/B_{\mathrm{LF}}^{\gamma}$, with $\gamma = 0.5$ for $x=0.08$.}
\end{figure}

As a next step, one can try to establish the nature of the spin-spin correlation function and, from this, the nature of the dynamic component of the relaxation. By using the same method adopted in spin-glass systems, in Fig.~\ref{fig:fig11} we display the time-dependent asymmetry as a function of the scaled variable $t/B_{\mathrm{LF}}^\gamma$. To exclude from this analysis the effects of the static component of the local field at the muon site, we removed the short-time data (i.e., $t < 1$\,$\mu$s).
With a value of $\gamma = 0.5$ the $t$-dependent muon asymmetries obtained at different fields overlap over almost two orders of magnitude in $t/B_{LF}^\gamma$\cite{Keren1996,Keren2004}, except for very low field values, where the static local field is expected to dominate. Since the scaling exponent $\gamma$ is lower than one, it suggests that at the \musr\ time scale the spin-spin correlation function $q(t)$ is best approximated by a power law, rather than by a stretched exponential, hence suggesting a picture of \textit{cooperative critical spin fluctuations} rather than a distribution of local fluctuation rates. This finding seems in marked contrast with the field- and temperature-behavior of the stretched exponential parameter, $\beta < 1$, which hints at a broad inhomogeneous distribution of the local magnetic fluctuation rates. The scenario of an inhomogeneous sample is, nevertheless, very unlikely: (i) the presence of a peak in $\lambda_
L(T)$ at about 30\,K, that should be absent in a fully disordered system; (ii) at $T = 30$\,K not less than 70\% of the sample volume shows static ordered magnetism, and (iii) as suggested by the x-ray analysis and confirmed by a rather sharp superconducting transition (relative to this family), the $x = 0.08$ sample is expected to be fairly homogeneous. Therefore, as in the case of CeFePO,\cite{Lausberg2012} the sizable dynamic component evidenced in our sample most likely reflects a cooperative scenario.\\
Interestingly, the critical slowing down of spin fluctuations occurs at sufficiently low temperatures, i.e., at $T <20$\,K, where both the static magnetism and the superconductivity occur over the whole sample volume. Unfortunately, we have no means of distinguishing whether the dynamic component can be related to the superconducting state or, most likely, it reflects a glassy magnetic phase.\\
Finally, we remark that the disordered magnetic phase seems a constant feature of the M-SC crossover region of the iron-based superconductors that was observed, for instance, also in iron chalcogenides at intermediate Se-Te substitutions.\cite{Lamura2013}
This type of short-range (disordered) magnetism could be driven by orbital nematic correlations induced by the tendency towards ordering of the iron 3$d$ orbitals.\cite{Lv2011,Martinelli2012,Fernandes2014}
To address this fundamental issue and unveil the true nature of the disordered magnetic phase, a systematic study of the structural and of the in-plane transport properties vs.\ temperature on high quality \nfao\ single crystals should be seriously considered.
\section{\label{sec:concl}Conclusion}
In this work we reported on the evolution with F-doping of the magnetic and superconducting properties of the Nd1111 family. We have shown that its electronic phase diagram presents similar features to other compounds of the Ln1111 family (Ln = La, Sm, Ce). In particular, at low doping ($x < 0.06$) a long-range antiferromagnetic order sets in for $T<T_{\mathrm{N}}$. At intermediate doping ($x \approx 0.08$) a short-range magnetic order develops which, at low temperature, coexists with bulk superconductivity on a nanometer length scale. Finally, at high F-doping values ($x \approx 0.25$), only the superconductivity survives, with no relevant traces of any magnetic order.

Contrary to the other Ln1111 cases, in the intermediate F-doping regime of Nd1111, we find clear evidence of fluctuating magnetism on the \musr\ time scale. This \textit{dynamic} magnetic component persists down to low temperature, until a short-range \textit{static} magnetism and superconductivity develop over the whole sample volume. This peculiarity of Nd1111 seems to be related to the glassy nature of its magnetically ordered phase at intermediate temperatures, i.e., just below the average N\'eel temperature.
\begin{acknowledgments}
We are grateful to A.\ Amato for instrumental assistance and thank P.\ Carretta for illuminating discussions and the careful reading of the manuscript. T.S.\ and N.Z.\ acknowledge support by the Schwei\-ze\-ri\-sche Na\-ti\-o\-nal\-fonds zur F\"or\-de\-rung der Wis\-sen\-schaft\-lich\-en For\-schung (SNF) and the NCCR research pool MaNEP of SNF.  We acknowledge also support by the FP7 European project SUPER-IRON (Grant Agreement No.\ 283204).
\end{acknowledgments}


%

\end{document}